\documentclass[preprint]{elsarticle}
\pdfoutput=1
\DeclareGraphicsExtensions{.pdf}
\usepackage{atlasphysics}
\usepackage{amssymb}
\usepackage{hyperref}

\usepackage[nodots]{numcompress}

\usepackage{lineno}

\newcommand{\PbPb}{lead-lead }

\newcommand{\sumet}{\mathrm{FCal} \; \Sigma E_{T}}
\journal{Physics Letters B}
\begin{document}

\begin{frontmatter}

\title{
\vspace{-0.5cm} \flushright{\normalsize{CERN-PH-EP-2011-124}} \\ \center{
Measurement of the pseudorapidity and transverse momentum dependence of the elliptic flow of charged particles in lead-lead collisions at $\sqrt{s_{_\mathrm{NN}}}=2.76$ TeV with the ATLAS detector}
}

\author{The ATLAS Collaboration}

\address{}

\begin{abstract}
This paper describes the measurement of elliptic flow of charged particles in lead-lead 
collisions at $\sqrt{s_{_\mathrm{NN}}}=2.76$ TeV using the ATLAS detector at the Large Hadron Collider (LHC). 
The results are based on an integrated luminosity of approximately 7 $\mu$b$^{-1}$.
Elliptic flow is measured over a wide region in pseudorapidity, $|\eta|<2.5$, and over a broad range in transverse momentum, $0.5 < \pT < 20$~GeV. 
The elliptic flow parameter $v_2$ is obtained by 
correlating individual tracks with the event plane measured 
using energy deposited in the forward calorimeters.
As a function of transverse momentum, $v_2(\pT)$ reaches a maximum at $\pT$ 
of about 3 GeV, then decreases and becomes weakly dependent on $\pT$ 
above 7--8 GeV.  
Over the measured pseudorapidity region, $v_2$ is found to be only weakly dependent on \eta, with less variation than observed at lower beam energies.
The results are discussed in the context of previous measurements at lower collision energies, as well as recent results from the LHC.

\end{abstract}


\end{frontmatter}

\section{Introduction}
\label{intro}
The measurement of collective phenomena in nuclear collisions at high energies 
has been a subject of intensive theoretical and experimental studies.
Anisotropic flow, which manifests itself as a large anisotropy in the 
event-by-event azimuthal angle distribution of produced particles, is generally understood to be a consequence 
of the spatial anisotropy of the initial energy deposition from nucleon-nucleon collisions in the overlap of the colliding nuclei.  
Anisotropies in the initial energy density are converted 
into final state momentum anisotropies via strong rescattering processes which 
induce pressure gradients, following the laws of relativistic hydrodynamics.
Consequently,
azimuthal anisotropies are sensitive to the initial
state and its subsequent dynamical evolution.  

Anisotropic flow is commonly studied 
by measuring the Fourier coefficients ($v_n$) of the azimuthal angle 
distributions of the emitted particles. 
The second harmonic, $v_2$, referred to as ``elliptic flow'', is the 
most extensively studied as it most directly relates the anisotropic shape of the
overlap of the colliding nuclei to a corresponding anisotropy of the outgoing momentum
distribution (for a review, see Ref.~\cite{Voloshin:2008dg}).
Elliptic flow
has been measured over a wide range of energies, collision 
systems, and collision centralities
by all of the RHIC heavy ion experiments~\cite{Arsene:2004fa,Ackermann:2000tr,Back:2004je,Adcox:2004mh} 
and several experiments at lower energies (see Ref.~\cite{Voloshin:2008dg}). 
Predictions for $v_2$ at the LHC energy varied widely, covering all possibilities from a strong rise, no change, or even a decrease 
of $v_2$~\cite{Abreu:2007kv} relative to lower energy collisions. 
Measurements of $v_2$ for inclusive charged-particles from the ALICE experiment \cite{alicepaper} indicate that,
integrated over $\pT$,
it increases by about 30\% from RHIC to LHC energies.  
However, ALICE also observed that $v_2(\pT)$ for inclusive charged particles was identical with RHIC results for the same
collision centrality (or impact parameter) 
up to $\pT = 4$ GeV.  
This implies that the observed rise is driven primarily by an increase in the average transverse momentum
with the higher collision energy.

In this letter, we present a measurement of the elliptic flow of charged particles in 
lead-lead collisions 
at $\sqrt{s_{_\mathrm{NN}}}=2.76$ TeV with the ATLAS detector at the LHC.
The elliptic flow is measured in the pseudorapidity region $|\eta|<2.5$ over the full azimuthal range
$0 < \phi < 2\pi$, for transverse momenta\footnote{ATLAS uses a right-handed coordinate system with its origin at the nominal interaction point (IP) in the centre of the detector and the $z$-axis along the beam pipe. The $x$-axis points from the IP to the centre of the LHC ring, and the $y$ axis points upward. Cylindrical coordinates $(r,\phi)$ are used in the transverse plane, $\phi$ being the azimuthal angle around the beam pipe. The pseudorapidity is defined in terms of the polar angle $\theta$ as $\eta=-\ln\tan(\theta/2)$.
Transverse momentum and energy are defined as $p_\mathrm{T} = p \sin \theta$ and $E_\mathrm{T} = E \sin \theta$, respectively.} $0.5 < p_\mathrm{T} < 20$ GeV.
This allows stringent tests of the applicability of hydrodynamics in the LHC energy regime, 
and provides information on 
the transition between low $\pT$, where hydrodynamics is expected to dominate, and higher $\pT$, 
where particle production is expected to stem from the fragmentation of jets modified by the hot, dense medium~\cite{Aad:2010bu}.

\section{ATLAS detector and trigger}
\label{detector}
The ATLAS detector~\cite{Aad:2008zzm} is well suited for measurements of azimuthal anisotropies over a large pseudorapidity range.  The relevant detectors for this analysis are the 
inner detector (ID) and forward calorimeter (FCal).
The ID
is contained within the 2~T field of a superconducting solenoid magnet,
and measures the trajectories
of charged particles in the pseudorapidity region $|\eta|<2.5$
and over the full azimuthal range.
The precision silicon tracking detectors consist of pixel detectors 
(Pixel) and a semiconductor microstrip tracker (SCT).  In the ``barrel'' 
region, these are arranged on cylindrical layers surrounding the beam pipe, while in 
the ``endcap'' regions they are mounted on disks perpendicular to the beam axis. 
A charged particle typically traverses three layers of the Pixel detector and 
four double-sided layers of the SCT.  The silicon detectors are surrounded by
a transition radiation tracker (TRT), composed of drift tubes and covering up to 
$|\eta|=2$.

The FCal covers a pseudorapidity range $ 3.2 <|\eta| < 4.9$.  
It uses tungsten and copper absorbers with liquid argon as the active medium,
with a total thickness of about 10 interaction lengths. 
This analysis uses the energy deposition in the entire FCal for the centrality
determination, while for the reaction plane measurement
only the energy deposition in the first sampling layer of the FCal (Layer 1) is used, 
as doing this was found to minimize the effect of fluctuations on the reaction plane measurement.

The trigger system has three stages, 
the first of which (Level-1) is hardware-based, while the 
later stages (Level-2 and Event Filter~\cite{Aad:2008zzm}) are based on software algorithms.
The minimum-bias Level-1 trigger used for this analysis 
requires signals in either the two sets of
minimum-bias trigger scintillator (MBTS) counters,
covering $2.1<|\eta|<3.9$ on each side of the experiment,
or the two zero-degree calorimeters (ZDC), each
positioned at $|z|=140$~m relative to the centre of the detector, detecting
neutrons and photons with $|\eta|>8.3$.
The ZDC Level-1 trigger thresholds were set just below the single neutron peak on each side.
The MBTS trigger was configured to require at least
one hit above threshold from each side of the detector.
A Level-2
timing requirement on signals from the MBTS
was then imposed to remove beam backgrounds, while the ZDC had no
further requirements beyond the Level-1 decision.
The Event Filter was not needed for the minimum-bias triggering and was run in pass-through mode.
\section{Event selection and reconstruction}
\label{data}

The \PbPb data set analysed here corresponds to an integrated luminosity of 
approximately $L_{\mathrm{int}} = 7$ $\mu\mathrm{b}^{-1}$.
Three main event selection requirements were applied offline 
to reject both non-collision backgrounds and Coulomb processes, in 
particular highly-inelastic photonuclear events.
First, an offline event selection required a time difference $|\Delta t| < 3$ ns 
between the positive and negative $\eta$ MBTS counters as well as a reconstructed vertex 
in order to suppress non-collision backgrounds. 
Second, a coincidence of the ZDCs at forward and backward pseudorapidities was required in order to
reject a variety of background processes, while maintaining high efficiency for non-Coulomb processes.
Finally, in this analysis only events with a vertex with $|z_{\mathrm{vtx}}|<10$ cm were used.
Simulations show the vertex algorithm to be
essentially 100\% efficient for the event sample considered here. 
Pile-up events, defined as additional minimum bias
events in the same bunch crossing, are expected to be present at the $10^{-4}$ level and so are negligible.
In total, approximately $4 \times 10^7$ events passed the selection criteria.

Tracks were reconstructed within the full acceptance of the inner detector.  
To improve the reliability of the ID track reconstruction in the
tracking environment in heavy ion collisions, the track quality requirements are more 
stringent than those defined for proton-proton collisions~\cite{ATLASminb}.
Tracks are required to have at least eight hits in the SCT, at least two Pixel hits and 
a hit in the Pixel layer closest to the interaction point.
A track must have no missing Pixel hits and at most one missing SCT hit, where such hits are expected.
Finally, the transverse and longitudinal impact parameters with respect to the 
vertex ($|d_0|$ and $|z_0 \sin \theta|$) were each required to be less than 1 mm.
These additional requirements were made to improve the purity of the track sample. 
The inefficiency of this selection is driven by the loss due to hadronic interactions 
in the detector material, which increases with $|\eta|$~\cite{ATLASminb}.
This results in an additional inefficiency of approximately 15\% at $|\eta| > 1$ 
compared to the central region of the detector.

However, the results shown here are found to be insensitive to the absolute tracking 
efficiency (discussed below), and the effect of the efficiency decrease
at high $|\eta|$ is minimized
when measurements are performed in limited transverse momentum and pseudorapidity intervals.

The tracking performance has been studied in detail by comparing data to Monte Carlo simulations
based on the HIJING event generator \cite{hijing} and a full GEANT4 \cite{geant4} simulation of the detector~\cite{ATLASSim}. 
In general the simulated distributions of the number of Pixel and SCT hits on tracks describe the 
data well, particularly after 
reweighting the simulated momentum distribution to account for the differences
in the charged particle spectrum reconstructed in data and HIJING.
Monte Carlo calculations show that the 
tracking efficiency for charged hadrons in this analysis 
is about 72\% near $\eta=0$ in central
collisions, lower than in proton-proton collisions due to the more stringent 
requirements and the higher occupancy in the SCT. 
Fake tracks from random combinations of hits are generally negligible, e.g.
reaching only 0.1\% in $|\eta|<0.3$ for the highest multiplicity collisions, although the rate of fake
tracks increases slightly with increasing $|\eta|$.

\section{Data analysis} 
\label{analysis}

\begin{figure}[t]
\begin{center}
\includegraphics[width=0.9\textwidth]{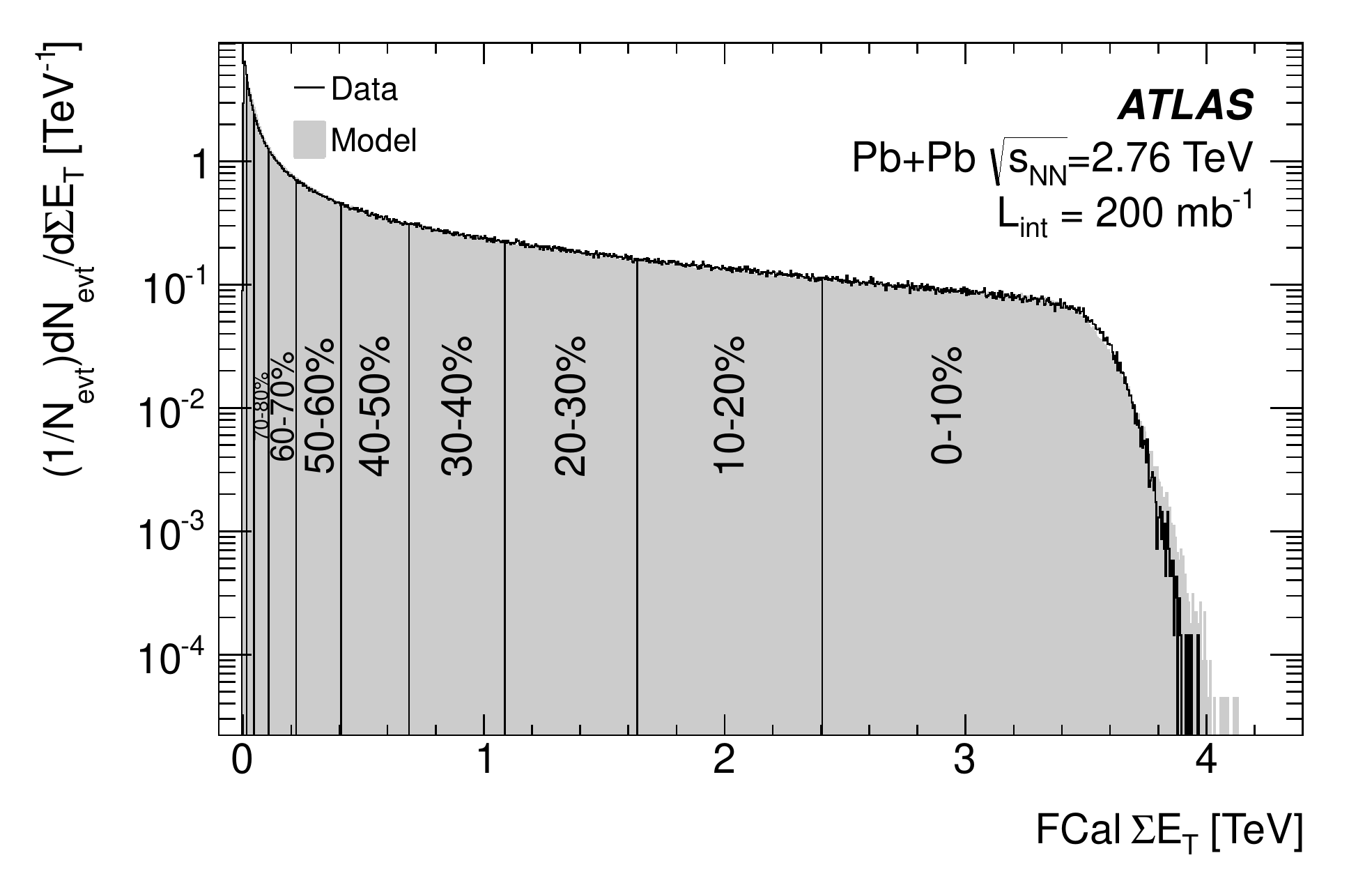}
\caption[fcal et distributions]{\label{fig:fcalcent}
Measured $\sumet$ distribution divided into 10\% centrality intervals (black).  Proton-proton
data at $\sqrt{s}=2.76$ TeV, convolved with a Glauber Monte Carlo calculation with $x=0.088$ (grey), 
as described in the text.}
\end{center}
\end{figure}

In order to systematically select various geometries of the initial state,
the data were analysed in centrality intervals defined by selections on $\sumet$, the total 
transverse energy deposited in the FCal (always stated at the electromagnetic energy scale,
which does not correct for the response of the calorimeter to hadrons).
These intervals are expressed in percentiles of the total inelastic non-Coulomb \PbPb cross section 
(0--10\%, 10--20\%,..., 70--80\%) with the most central interval (0-10\%) corresponding to 
the 10\% of events with the largest $\sumet$. 
The measured $\sumet$ distribution for a subset 
of the data (with $L_{\mathrm{int}}$ approximately 200 mb$^{-1}$),
taken with a less restrictive primary trigger than used for the bulk 
of the data and used for the calibration procedure described below, is shown 
divided into centrality intervals in Fig.~\ref{fig:fcalcent}.

To establish the fraction $f$ of the total non-Coulomb inelastic
cross section selected by our
trigger and event selection criteria, we have performed a convolution of
$\sumet$ distributions measured in proton-proton data at 
$\sqrt{s}=2.76$ TeV with a full Monte Carlo Glauber calculation~\cite{Miller:2007ri}. 
The calculation assumes the number of effective proton-proton collisions per \PbPb event, $N$, scales according to
the ``two-component model'' (from e.g. Ref~\cite{Kharzeev:2000ph}).
This model combines the number of participants 
($N_{\mathrm{part}}$, 
the number of nucleons which interact inelastically at least once) and
the number of binary collisions ($N_{\mathrm{coll}}$) as 
$N = (1-x)\frac{N_{\mathrm{part}}}{2} + x N_{\mathrm{coll}}$.
In this approach, the only free parameter is $x$, which controls the relative
contribution of $N_{\mathrm{part}}$ and $N_{\mathrm{coll}}$.
The best description of the data is found 
to be for $x = 0.088$.  
The value of $f$ and its uncertainty is estimated by
systematically varying the effect
of trigger and event selection inefficiencies as well as backgrounds
in the most peripheral $\sumet$ interval to achieve the best agreement between
the measured and simulated distributions.
Using this analysis of the $\sumet$ distribution, the fraction of the 
total cross section sampled by the
trigger and event selection has been estimated to be 98\%, 
with an uncertainty of 2\%. 
This is similar  
to estimates given in a previous ATLAS publication~\cite{:2010px}.
The $\sumet$ ranges defined from this subsample have been found to be stable for the full data set,
both by counting the number of events and by measuring the average number of reconstructed tracks
in each interval.
The 20\% of events with the smallest $\sumet$ are not included
in this analysis, due to the relatively large uncertainties in determining
the appropriate selection criteria. 

The final state momentum anisotropy can be quantified by studying the Fourier
decomposition of the azimuthal angle distribution \cite{Poskanzer}:
\begin{equation}
E \frac{d^3N}{dp^3}= \frac{1}{\pT} \frac{d^3 N}{d\phi d\pT dy} = \frac{1}{2\pi \pT}\frac{E}{p}\frac{d^2N}{d\pT d\eta}  \left( 1+2\sum_{n=1}^{\infty} v_n \cos{[n(\phi-\Psi_{n})]} \right) ,
\label{eq:fourier}
\end{equation}
where $y$, $\pT$ and $\phi$ are the rapidity, transverse momentum, and azimuthal angle of 
final-state charged particle tracks and $\Psi_{n}$ denotes 
the azimuthal angle of the $n$-th order reaction plane.  
In more peripheral events, $\Psi_2$ is close to $\Phi_{RP}$, the reaction plane angle,
defined by the impact parameter ($\vec{b}$, the vector separation of the barycentres of the two nuclei)
and the beam axis ($z$). 
In more central events, $\Psi_2$ primarily reflects fluctuations in the initial-state configurations of colliding nucleons.
This analysis was confined to the second
Fourier coefficient ($n=2$), $v_2 \equiv \left<\cos{[2(\phi-\Phi_{RP})]}\right> $, 
where angular brackets denote an average first over particles within each event relative to the eventwise
reaction plane, and then over events. 

In this analysis, 
the $n=2$ event plane is determined from the data on an event-by-event basis,
according to the scheme outlined in Ref.~\cite{Poskanzer}:
\begin{equation}
\label{eq:eptracks}
\Psi_2 = \frac{1}{2}\tan^{-1}\left( \frac{\sum E^{\mathrm{tower}}_{\mathrm{T},i} w_i\sin(2\phi_i)}
{\sum E^{\mathrm{tower}}_{\mathrm{T},i} w_i\cos(2\phi_i)}\right),
\end{equation}
\noindent
where sums run over tower transverse energies $E_\mathrm{T}^{\mathrm{tower}}$ as 
measured in the first sampling layer of the forward calorimeters, with each tower covering
$\Delta \eta \times \Delta \phi = 0.1 \times 0.1$. 
The tower weights, $w_i = w_i(\phi_i,\eta_i)$, are used to correct for local variations in detector response.
They are calculated in narrow $\Delta\eta$ slices ($\Delta \eta = 0.1$) over the full FCal $\eta$ range in such 
a way as to remove structures in the uncorrected $\phi$ distributions of $E_\mathrm{T}^{\mathrm{tower}}$ in every $\Delta\eta$ slice. 
The final results of this analysis are found to be insensitive to the 
weighting, and results obtained with all $w_i=1$ were consistent with those reported here,
and well within the systematic uncertainties estimated below.

\begin{figure}[t]
\begin{center}
\includegraphics[width=0.7\textwidth]{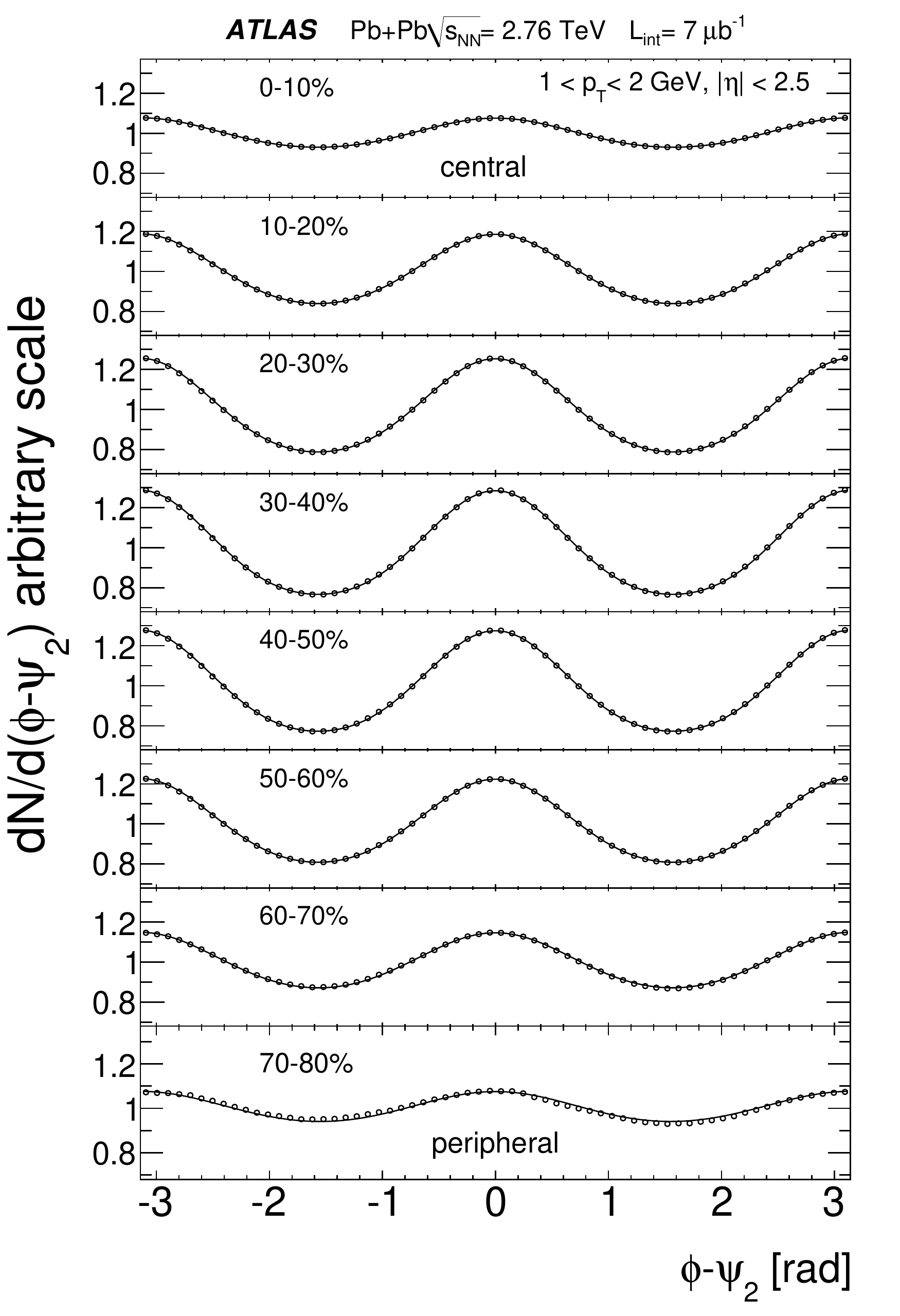}
\caption[phi minus psi]{\label{fig:eptrcorr}
Distribution of the azimuthal angle of individual tracks relative to the measured event plane, in eight centrality intervals.
These distributions are meant to illustrate the observed correlation relative to the event plane, and are not used
in the quantitative estimates of $v_2$.
The curve is a fit to $1+\sum_n 2v_n \cos(n\phi)$ up to $n=6$.
} 
\end{center}
\end{figure}

The correlation of individual track azimuthal angles with the estimated event plane is shown in Fig.~\ref{fig:eptrcorr}
for tracks with $\pT = 1-2$ GeV.  
There is a clear sinusoidal modulation at all centralities.  The modulation is largest in 
the 20--50\% centrality intervals, and decreases for the more central and peripheral events.  
In the centrality intervals where the correlation is strongest, 
the correlation does not follow 
a perfect $1+\alpha \cos(2\phi)$ form, indicating significant
contributions from higher order harmonics.  
However, in this letter we rely on the orthogonality 
of the Fourier expansion and do not extract the other coefficients.  
To verify that this does not bias the measurement, we have extracted
$v_2$ from a fit containing all Fourier components $v_n$ up to $n=6$, and found
$v_2$ values consistent with the results extracted below.  
The odd amplitudes are found to be consistent with zero, 
as expected when measuring odd harmonic functions relative to $\Psi_2$~\cite{Poskanzer}.

\begin{figure}[t]
\begin{center}
\includegraphics[width=0.48\textwidth]{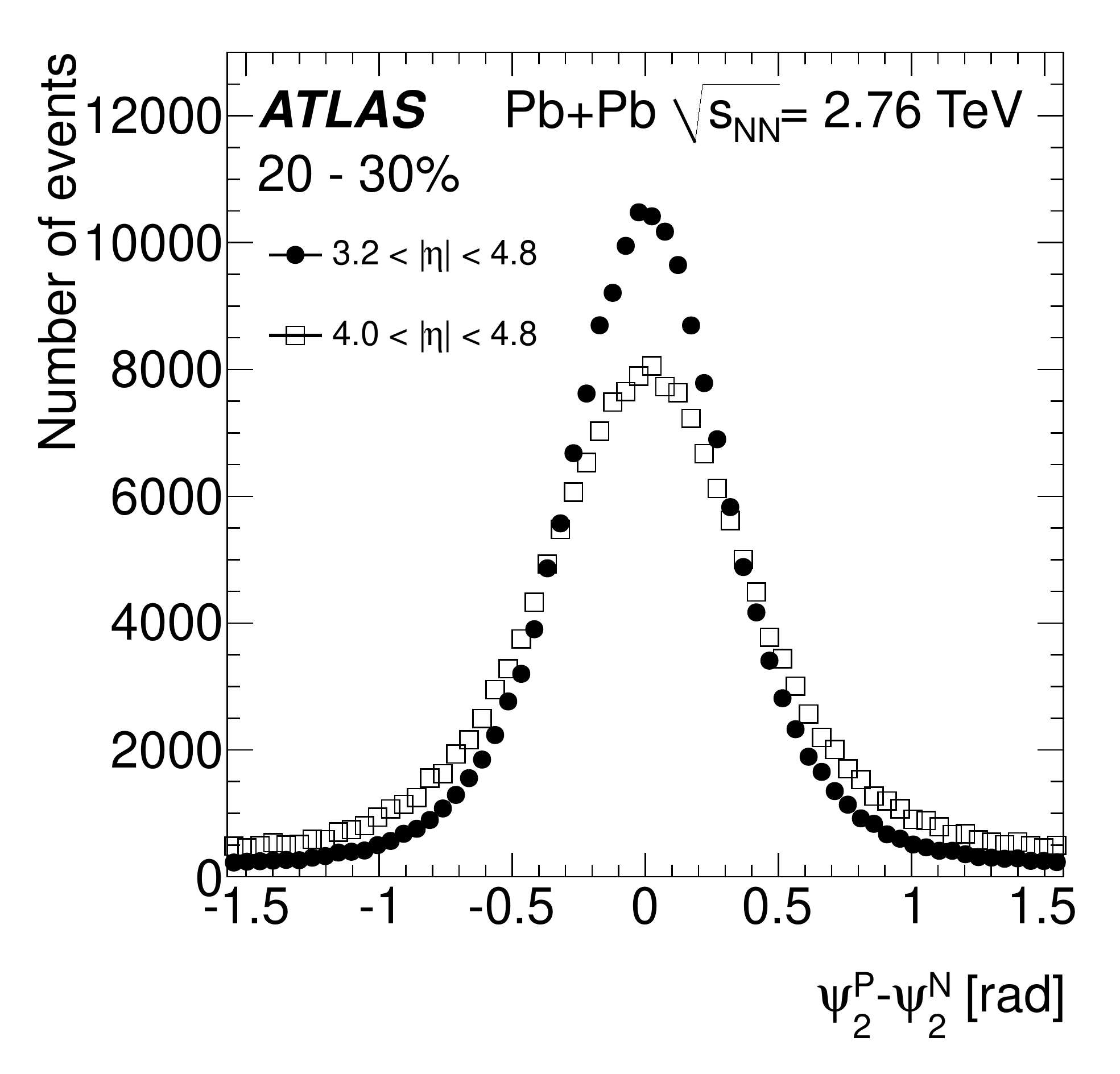}
\includegraphics[width=0.48\textwidth]{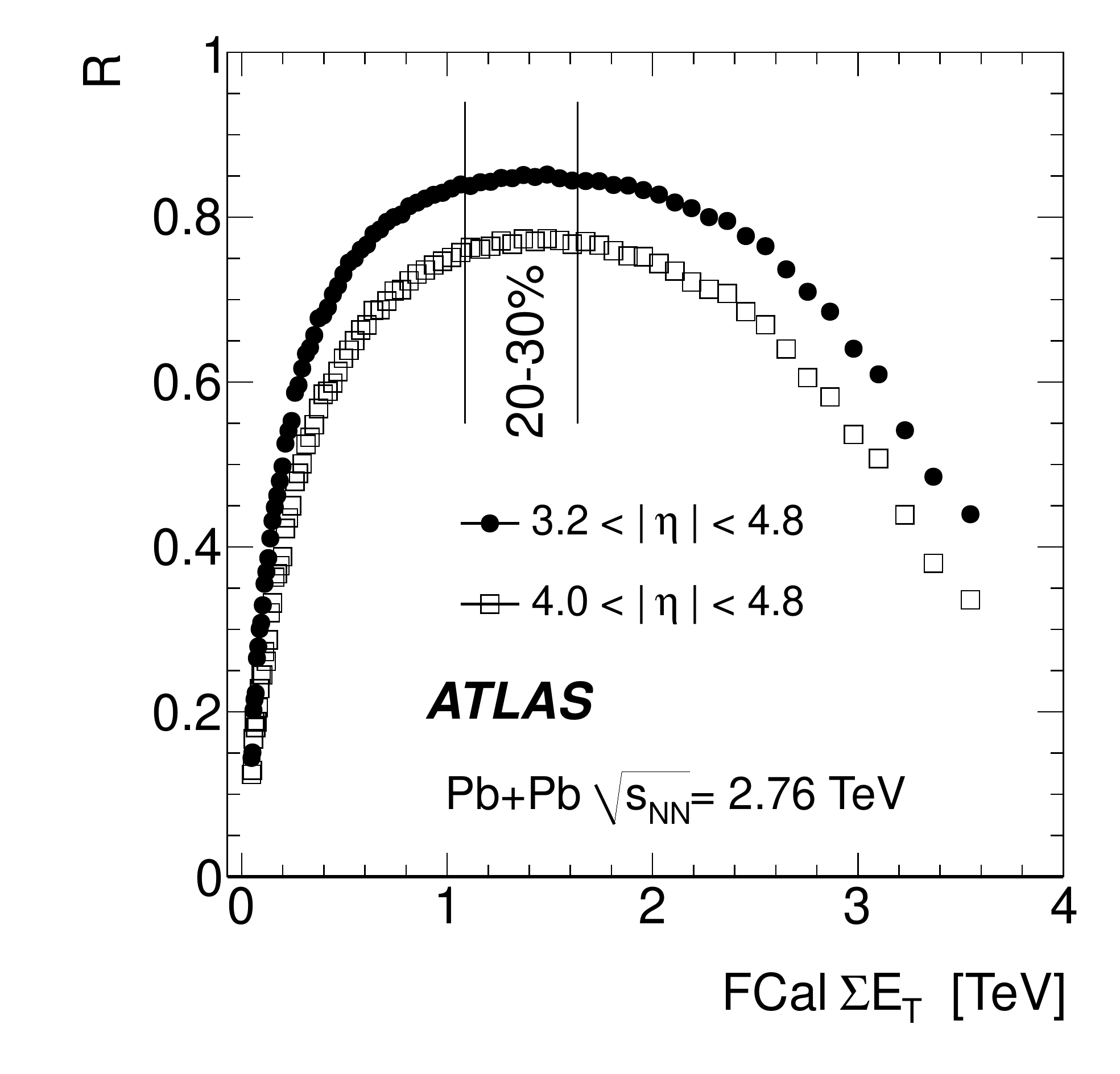}
\caption[res comparison fcl0 trk]{\label{fig:epcorr}
(left) 
Distribution of the difference between the event planes at positive and negative $\eta$ 
obtained using Layer 1 FCal towers, both with
full and half acceptance.
(right) $\sumet$ dependence of the resolution correction for event planes from Layer 1 FCal towers 
in full acceptance (full symbols) and half acceptance (open symbols).

}
\end{center}
\end{figure}

The measured values of $v_2$ are generally underestimated because of the 
finite experimental resolution in extracting the event plane angle.
The event plane resolution correction factor, $R$, was obtained using the subevent technique, also described 
in Ref.~\cite{Poskanzer}. 
Two ``subevents'' are defined in each event, one each in the forward and backward $\eta$ directions. 
For the measurement of the event plane using the FCal, the first sampling layer on the 
positive $\eta$ side was selected as subevent ``$P$'', with a corresponding subevent ``$N$''
formed for negative $\eta$. 
The resolution correction for the event plane measured by each subevent
was calculated as a function of $\sumet$ according to the formula
\begin{equation}
\label{eq:rescorrdata}
R(\Sigma E_\mathrm{T}) = \sqrt{\left< \cos[2(\Psi_2^N-\Psi_2^P)] \right>},
\end{equation}
where angular brackets denote an average over all events in a $\sumet$ interval.
The left panel of Fig.~\ref{fig:epcorr} shows the distribution of the difference $\Psi_2^P-\Psi_2^N$.
The right panel shows the $\sumet$ dependence of the resolution correction for the event plane determined
using the full FCal Layer 1 as well as a reduced-acceptance version used in the systematic studies discussed below.

The final, resolution-corrected, $v_2$ is calculated in intervals of centrality, $\eta$ and $\pT$ as 
\begin{equation}
\label{eq:v2datarun}
v_2(\eta,\pT) = \frac{1}{N_{\mathrm{tot}}^{\mathrm{trk}}}\sum_{j}^{\mathrm{events}}\frac{1}{R(\Sigma E_\mathrm{T})}\sum_{i}^{\mathrm{tracks}}{c_{i}\cos{[2(\phi_i-\Psi^{N/P}_{2,j})]}},
\end{equation} 
\noindent
where $N_{\mathrm{tot}}^{\mathrm{trk}}$ denotes the total number of the reconstructed tracks in a given centrality, $\eta$ and $\pT$ range, and the $c_i$ are weights similar to
the $w_i$ for tracks, designed to flatten the $\phi$ distribution in a
small $\Delta\eta$ slice.
For $\Psi^{N/P}_{2,j}$ (the event plane for event $j$) we take the event plane 
measured in the opposite $\eta$ hemisphere (i.e. ``$P$'' at positive $\eta$, or ``$N$'' at negative $\eta$) 
to each track with azimuthal angle $\phi_i$. 
Using the track in the opposite hemisphere maximizes the pseudorapidity 
gap between the 
reaction plane estimate and the $v_2$ estimate ($|\Delta \eta|>3.2$), 
minimizing potential non-flow correlations between them.

The systematic uncertainty on $v_2$ as a function of
$\pT$, $\eta$ and centrality was evaluated by varying several different aspects of the analysis procedure. 
\begin{itemize}
\item The resolution correction was changed by limiting the FCal acceptance to a smaller range in pseudorapidity.
\item Tighter tracking requirements were applied (both $|d_0|$ and $|z_0 \sin \theta|$ less than 0.5 mm, instead of the nominal 1 mm requirement).
\item Results were compared using negatively and positively charged tracks.
\item Results were compared between $v_2$ measured at positive and negative pseudorapidities.
\item Results were studied as a function of time during the heavy ion run.
\end{itemize}

Additional sources of systematic uncertainties were examined, including the following:
Deviations from zero of $\langle \sin(2[\phi - \Psi_2]) \rangle$, which are sensitive to residual biases in the reaction plane determination and detector non-uniformities, were measured.
Monte Carlo studies were performed based on HIJING, with
a special procedure applied to the generated particle azimuthal angles so as 
to simulate elliptic flow (from Ref.~\cite{Poskanzer}), 
with a magnitude extrapolated from RHIC data.
Deviations from the flow induced at the generator level were obtained by applying the same 
analysis procedure to the simulated data as with real data. 
The event plane determined from the reconstructed tracks was also investigated as an independent cross-check 
on the FCal reaction plane.
In this case, for the tracks with positive (negative) $\eta$ the event plane determined in 
the negative (positive) $\eta$ subevent was used.
The uncertainty in the fraction of the total inelastic cross section sampled by our trigger
and event selection criteria gives an overall scale uncertainty on $v_2$, ranging from 1\% in central events up to 5\% in peripheral events.

Deviations in individual contributions from the baseline results have been quantified as
relative systematic uncertainties on $v_2$ (in percent),
which are listed in Table~\ref{tab:systematics} for several
centrality and $\pT$ intervals, all for $|\eta|<1$.
The different components 
have been added in quadrature and expressed as $1\sigma$ point-to-point systematic uncertainties.
Note that the somewhat large increase in the scale of the uncertainties from moderate to high \pT\ can 
be partly attributed to the limited track statistics at high $\pT$.
It should also be pointed out that the systematic uncertainties only include those associated with
the measurements themselves; no attempt is made to disentangle the potential contributions
from non-flow effects, since their nature is not yet fully understood.

\begin{table}
  \begin{center}
\footnotesize
    \begin{tabular}{|p{4cm}||c|c|c||c|c|c||}
      \hline
   \multicolumn{1}{|c||} {  Centrality }               &  \multicolumn{3}{c||}{0-10\% }        &   \multicolumn{3}{c||}{40-50\%}     \\
      \hline
   \multicolumn{1}{|c||} { $p_{T}$ [GeV]}               &  0.8-0.9  &  2.4-2.7  &   8-10        &  0.8-0.9  &  2.4-2.7  &  8-10      \\ 
      \hline 
Smaller $\eta$ acceptance of event plane determination  &   0.6     &   1.2     &   5.7         &   0.7    &    0.7     &  2.0       \\
      \hline                                                                                  
Residual deviation from zero of sine terms              &   0.7     &   0.6     &   0.4         &   0.5    &    0.7     &  1.2       \\
      \hline                                                                                 
Varying tracking cuts                                  &    0.4     &   0.1     &   1.7         &   0.1    &  $<0.1$    &  0.2       \\
      \hline                                                                                 
Negative vs positive tracks                            &    0.5     &   0.3     &   3.3         &   0.3    &    0.1     &  1.6       \\
      \hline                                                                                 
Asymmetry with respect to $\eta$ reflection            &    0.1     &   0.1     &   0.2         &  $<0.1$  &    $<0.1$  &  0.1       \\
      \hline                                                                                
Time dependence                                  &   \multicolumn{3}{c||}{0.2}            &   \multicolumn{3}{c||}{0.2}        \\
      \hline                                                                                 
Monte Carlo reconstruction                             &   1.2      &   1.2     &   1.2         &   0.3    &     0.3    &  0.3       \\
      \hline                                                                                 
Total systematic error                                 &   1.6      &   1.9     &   6.9         &   1.0    &     1.0    &  2.9       \\
      \hline                                                                                  
    \end{tabular}
\caption[sources of systematic errors]{Principal systematic uncertainties (stated as a percentage of the value of $v_2$) on the $v_2$ measurement for three $\pT$ intervals and two centrality intervals, all for $|\eta|<1$.}
  \label{tab:systematics}
  \end{center}
\end{table}

\section{Results}
\label{results}
\begin{figure}[t]
\begin{center}
\includegraphics[angle=0,width=0.9\textwidth]{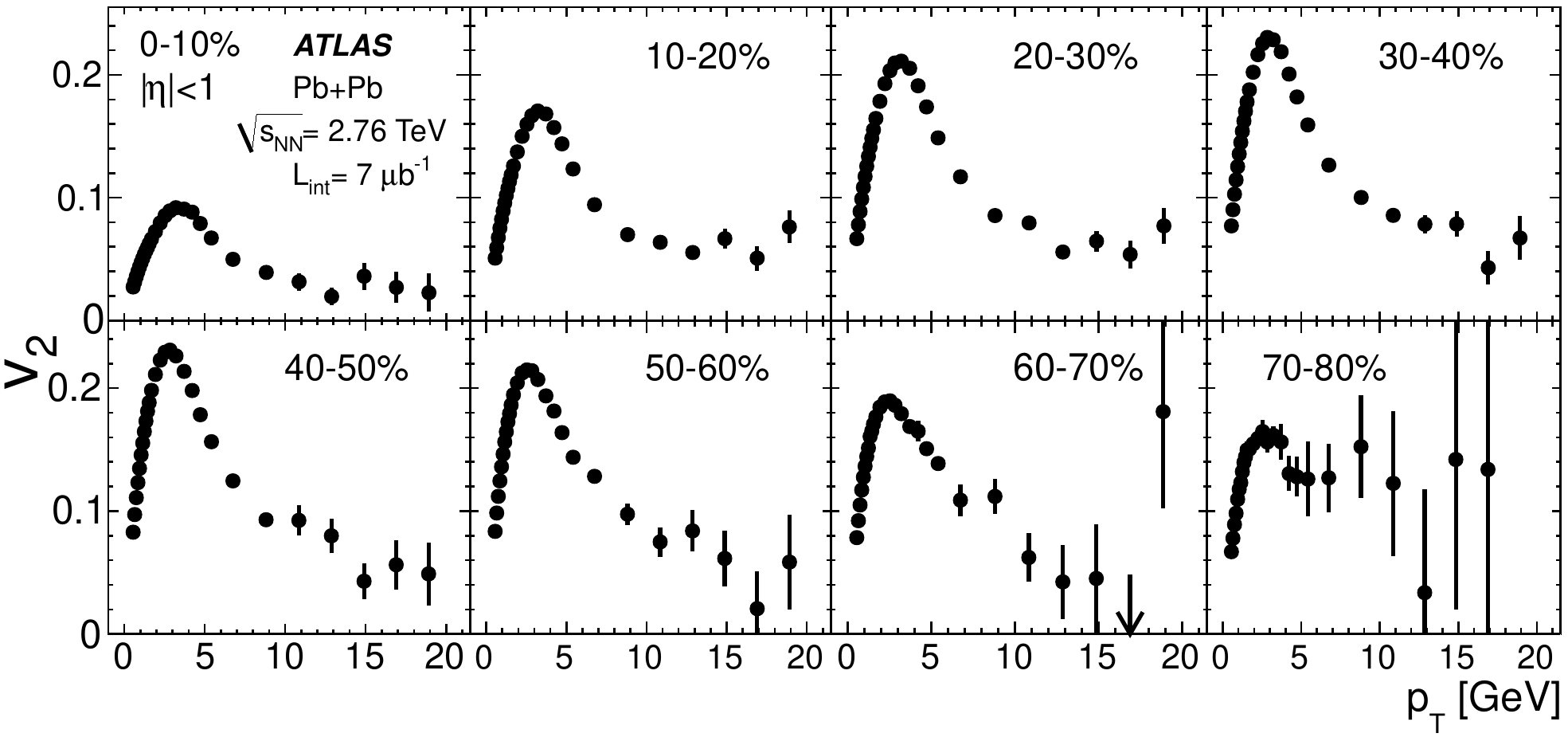}
\includegraphics[angle=0,width=0.9\textwidth]{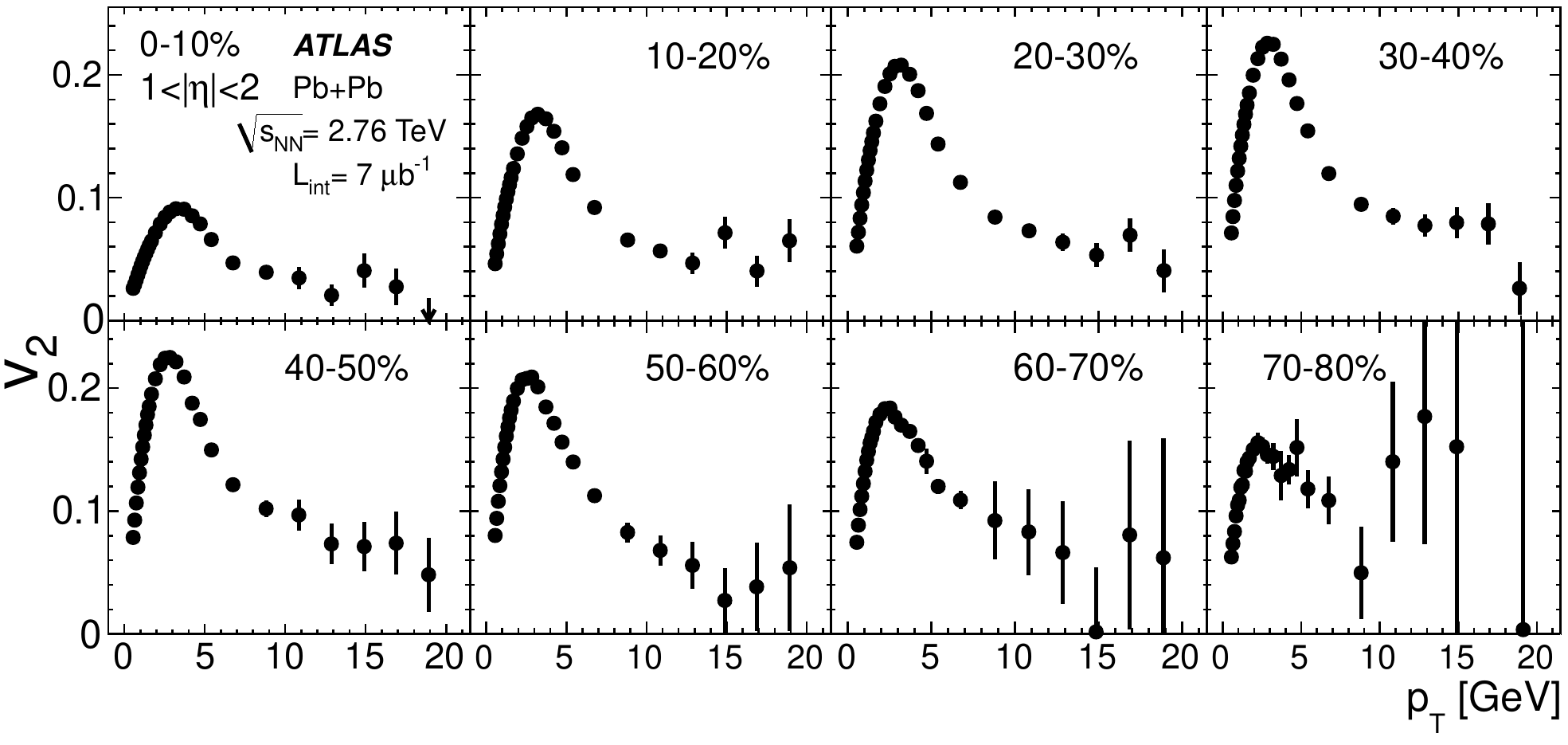}
\includegraphics[angle=0,width=0.9\textwidth]{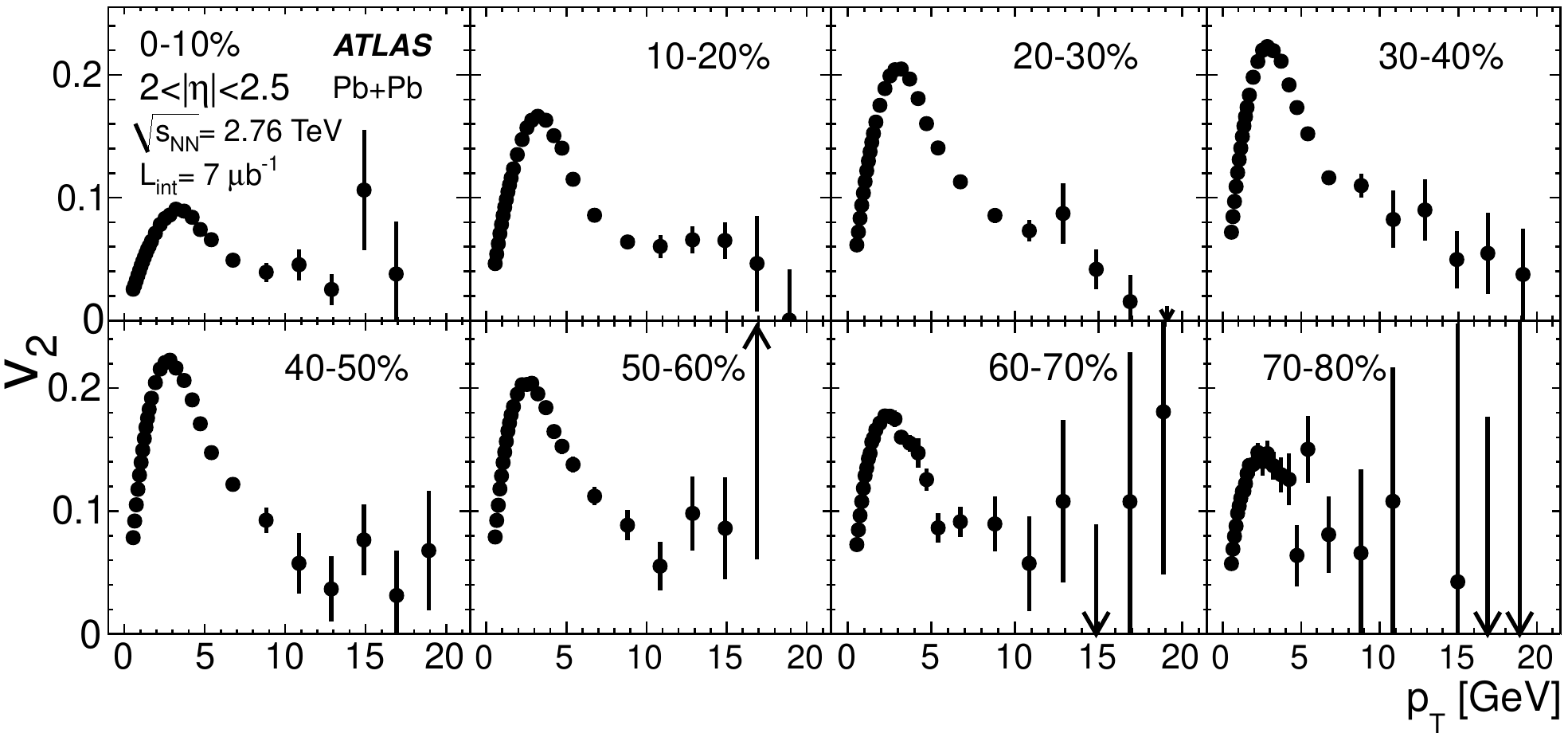}
\caption[pt dependence]{\label{fig:ptdep}
Elliptic flow $v_2(\pT)$ as a function of $\pT$
for eight 10\% centrality intervals, for $\pT$ from 0.5 to 20 GeV, and for
three ranges in pseudorapidity ($|\eta|<1$, $1<|\eta|<2$ and $2<|\eta|<2.5$).
Error bars show statistical and systematic uncertainties added in quadrature.
The arrows indicate where the value of $v_2$ does not fit within the chosen plot scale, due to large statistical fluctuations.}
\end{center}
\end{figure}

The top panel of Fig.~\ref{fig:ptdep} shows the $v_2$ dependence on $\pT$ for eight 10\% centrality intervals and for tracks with $|\eta|<1$. 
It is observed that all centrality intervals show a rapid rise in $v_2(\pT)$ up to $\pT =$ 3 GeV,
a decrease out to 7--8 GeV, and then a weak $\pT$-dependence beyond 9--10 GeV.
The same trends are also seen for $1<|\eta|<2$ (Fig.~\ref{fig:ptdep} middle) 
and $2<|\eta|<2.5$ (Fig.~\ref{fig:ptdep} bottom).

\begin{figure}[t]
\begin{center}
\includegraphics[angle=0,width=0.9\textwidth]{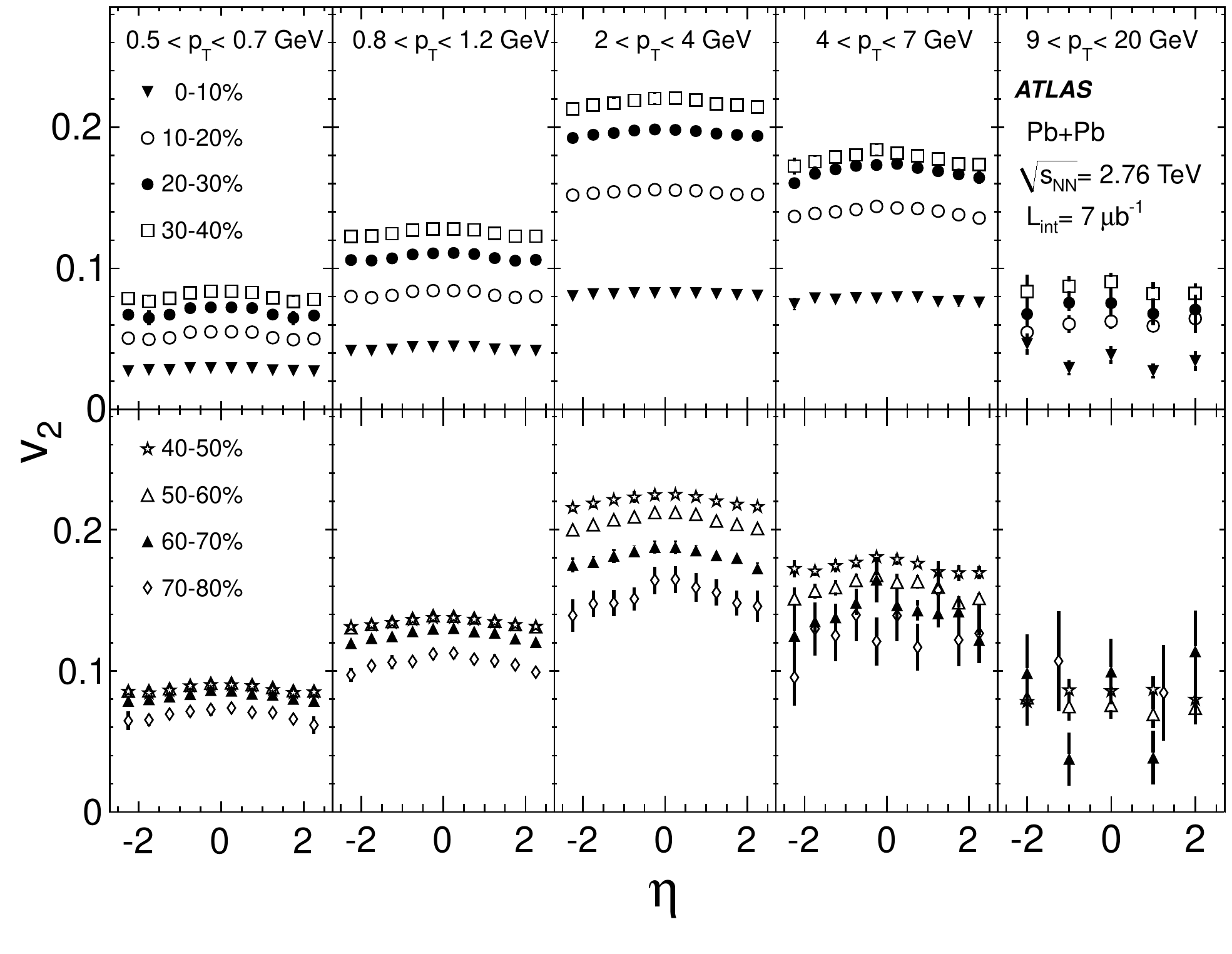}
\caption[pt dependence]{\label{fig:etadep}
Pseudorapidity dependence of $v_2(\pT,\eta)$ for $0.5 < \pT < 20$ GeV in five $\pT$ intervals and 10\% centrality intervals.  
Error bars show statistical and systematic uncertainties added in quadrature.
} 
\end{center}
\end{figure}

The pseudorapidity dependence of $v_2$ is shown in Fig.~\ref{fig:etadep}.
The top row shows the centrality and $\eta$ dependence of $v_2(\eta,\pT)$ for
five $\pT$ intervals, which characterize the trend shown in Fig.~\ref{fig:ptdep},
and the four most-central intervals.
The bottom row shows the same information for the four most-peripheral intervals.
It is observed that $v_2$ depends very weakly on $\eta$ over the measured 
pseudorapidity region.  In the two lowest $\pT$ intervals, below $1.2$ GeV,
$v_2$ drops
by about 5--10\% over the range $|\eta|= 0-2.4$. At higher transverse momenta,
a decrease on the order of few percent can be seen, although, due to the
large point-to-point errors, a flat $\eta$ dependence cannot be excluded. 
This is in contrast to the strong variation in $v_2(\eta)$ observed by the PHOBOS experiment
at $\sqrt{s_\mathrm{NN}}=200$ GeV~\cite{Back:2004mh}, which 
drops by approximately 30\% between $\eta=0$ and $\eta=2.5$.  

\begin{figure}[t]
\begin{center}
\includegraphics[angle=0,width=0.9\textwidth]{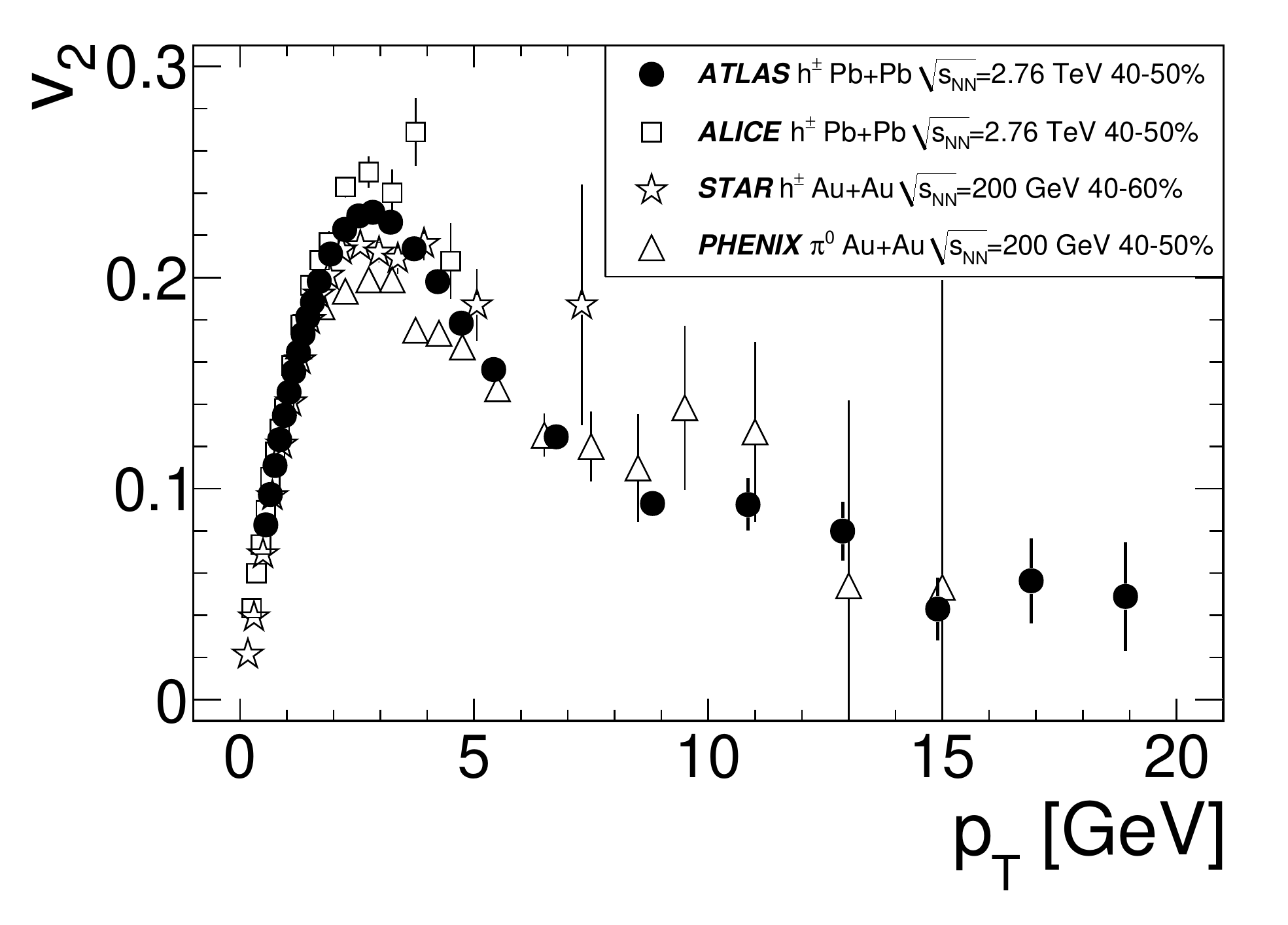}
\caption[pt dependence]{\label{fig:comp4050}
$v_2$ vs. $\pT$ at $|\eta|<1$ in the 40--50\% centrality interval, compared to previous
experimental data: ALICE $v_2\{2\}$\cite{alicepaper} for inclusive charged particles, PHENIX~\cite{Adare:2010sp} 
$v_2$ for identified $\pi^0$, and STAR data on $v_2\{2\}$ for inclusive 
charged particles for the 40--60\% interval~\cite{Adams:2004bi}.
} 
\end{center}
\end{figure}

Fig.~\ref{fig:comp4050} shows $v_2(\pT)$ for $|\eta|<1$ in the 40--50\% centrality interval compared to data 
from the LHC (ALICE, from Ref.~\cite{alicepaper}) as well as from RHIC (STAR~\cite{Adams:2004bi} 
and PHENIX~\cite{Adare:2010sp}) with a centre-of-mass energy a factor of nearly 14 lower.
The ALICE and STAR data are shown for the second cumulant $v_2\{2\}$, which gives results closest to the event-plane
method used in this analysis.  The PHENIX data are obtained 
with a similar method as ATLAS, but with $v_2$ measured only for identified $\pi^0$ hadrons,
detected through their two-photon decay mode.
It is observed that all of the data sets are quite similar as a function of $\pT$, 
both at lower $\pT$ (ALICE and STAR) and even at higher $\pT$,
within the limited statistical precision of the PHENIX data.
The observation of similar $v_2$ at low $\pT$ has been noted 
recently~\cite{alicepaper}, and has been reproduced using hydrodynamical
simulations assuming the same shear viscosity to entropy density ratio but initialized 
at a higher energy density. 
However, the similarities at high $\pT$ will require additional theoretical 
study to see if they are consistent with the differential energy loss of jets 
in the hot, dense medium.

\section{Conclusions}
\label{summary}
Elliptic flow measurements in \PbPb collisions at $\sqrt{s_{_\mathrm{NN}}}=2.76$ TeV 
obtained with the ATLAS detector are presented for an integrated luminosity of
approximately 7 $\mu$b$^{-1}$. 
These results represent the first measurement of $v_2$ 
over a broad range in $\eta$ and $\pT$ at the LHC energy. 
As a function of transverse momentum, at all $|\eta|$,  $v_2$ rises rapidly up to $\pT = 3$ GeV,
decreases somewhat less rapidly out to $\pT =$ 7--8 GeV, and then
varies weakly out to 20 GeV.
Over the measured pseudorapidity region, $|\eta|<2.5$, $v_2$ is found to be only weakly dependent on \eta, with less variation than observed at lower beam energies.
Comparison of the 40--50\% interval with lower energy data shows little change both at low and high $\pT$.
These results provide strong constraints on models which aim to describe the dynamical evolution 
of the system created in ultra-relativistic heavy ion collisions.   

\section{Acknowledgements}
We  thank CERN for the efficient commissioning and operation of the LHC during this initial heavy ion data taking period as well as the support staff from our institutions without whom ATLAS could not be operated efficiently.

We acknowledge the support of ANPCyT, Argentina; YerPhI, Armenia; ARC, Australia; BMWF, Austria; ANAS, Azerbaijan; SSTC, Belarus; CNPq and FAPESP, Brazil; NSERC, NRC and CFI, Canada; CERN; CONICYT, Chile; CAS, MOST and NSFC, China; COLCIENCIAS, Colombia; MSMT CR, MPO CR and VSC CR, Czech Republic; DNRF, DNSRC and Lundbeck Foundation, Denmark; ARTEMIS, European Union; IN2P3-CNRS, CEA-DSM/IRFU, France; GNAS, Georgia; BMBF, DFG, HGF, MPG and AvH Foundation, Germany; GSRT, Greece; ISF, MINERVA, GIF, DIP and Benoziyo Center, Israel; INFN, Italy; MEXT and JSPS, Japan; CNRST, Morocco; FOM and NWO, Netherlands; RCN, Norway;  MNiSW, Poland; GRICES and FCT, Portugal;  MERYS (MECTS), Romania;  MES of Russia and ROSATOM, Russian Federation; JINR; MSTD, Serbia; MSSR, Slovakia; ARRS and MVZT, Slovenia; DST/NRF, South Africa; MICINN, Spain; SRC and Wallenberg Foundation, Sweden; SER,  SNSF and Cantons of Bern and Geneva, Switzerland;  NSC, Taiwan; TAEK, Turkey; STFC, the Royal Society and Leverhulme Trust, United Kingdom; DOE and NSF, United States of America.  

The crucial computing support from all WLCG partners is acknowledged gratefully, in particular from CERN and the ATLAS Tier-1 facilities at TRIUMF (Canada), NDGF (Denmark, Norway, Sweden), CC-IN2P3 (France), KIT/GridKA (Germany), INFN-CNAF (Italy), NL-T1 (Netherlands), PIC (Spain), ASGC (Taiwan), RAL (UK) and BNL (USA) and in the Tier-2 facilities worldwide.

\clearpage
                                   
\label{app:ATLASColl}

\begin{flushleft}
{\Large The ATLAS Collaboration}

\bigskip

G.~Aad$^{\rm 48}$,
B.~Abbott$^{\rm 111}$,
J.~Abdallah$^{\rm 11}$,
A.A.~Abdelalim$^{\rm 49}$,
A.~Abdesselam$^{\rm 118}$,
O.~Abdinov$^{\rm 10}$,
B.~Abi$^{\rm 112}$,
M.~Abolins$^{\rm 88}$,
H.~Abramowicz$^{\rm 153}$,
H.~Abreu$^{\rm 115}$
% ...
\bigskip

% ...
$^{10}$ Institute of Physics, Azerbaijan Academy of Sciences, Baku, Azerbaijan\\
$^{11}$ Institut de F\'isica d'Altes Energies and Departament de F\'isica de la Universitat Aut\`onoma  de Barcelona and ICREA, Barcelona, Spain\\
% ...
$^{48}$ Fakult\"{a}t f\"{u}r Mathematik und Physik, Albert-Ludwigs-Universit\"{a}t, Freiburg i.Br., Germany\\
$^{49}$ Section de Physique, Universit\'e de Gen\`eve, Geneva, Switzerland\\
% ...
$^{88}$ Department of Physics and Astronomy, Michigan State University, East Lansing MI, United States of America\\
% ...
$^{111}$ Homer L. Dodge Department of Physics and Astronomy, University of Oklahoma, Norman OK, United States of America\\
$^{112}$ Department of Physics, Oklahoma State University, Stillwater OK, United States of America\\
% ...
$^{115}$ LAL, Univ. Paris-Sud and CNRS/IN2P3, Orsay, France\\
% ...
$^{118}$ Department of Physics, Oxford University, Oxford, United Kingdom\\
% ...
$^{153}$ Raymond and Beverly Sackler School of Physics and Astronomy, Tel Aviv University, Tel Aviv, Israel\\
% ...
\end{flushleft}


\end{document}